\documentstyle[11pt,epsfig,psfig]{article}
\def\be{\begin{eqnarray}}
\def\ee{\end{eqnarray}}
\def\MeV{\mbox{MeV}}
\def\fm{\mbox{fm}}
\def\roughly#1{\mathrel{\raise.3ex\hbox{$#1$\kern-.75em%
\lower1ex\hbox{$\sim$}}}}
\def\lsim{\roughly<}
\def\gsim{\roughly>}

\begin{document}
\begin{center}
%\begin{flushright}
%  \version
%\end{flushright}
\vskip 1cm {\Large \bf A Higher-Order Calculation of $np$ Scattering\\
 in Cut-Off Effective Field Theory}
\vskip 1.5cm {
{\large Chang Ho Hyun}$^{a,}$\footnote{e-mail: hch@zoo.snu.ac.kr}, 
{\large Tae-Sun Park}$^{b,}$\footnote{e-mail: park@alph02.triumf.ca} and
{\large  Dong-Pil Min}$^{a,c,}$\footnote{e-mail: dpmin@phya.snu.ac.kr} \\ 
\vskip 0.1cm 
$^a${\it Department of Physics, Seoul National
University, Seoul 151-742, Korea}} \\  $^b${\it Theory
Group, TRIUMF, 4004 Wesbrook Mall, Vancouver, B.C., Canada V6T 2A3
\\  $^c${\it Service de Physique Th\'{e}orique, CE
Saclay, 91191 Gif-sur-Yvette, France}}

\end{center}
\centerline{\bf Abstract} \vskip 0.1cm

We report a next-to-leading-order (NLO) chiral perturbation theory
calculation of the neutron-proton scattering cross section in the
${}^1S_0$ channel using a cut-off regularization. The inclusion of
two-pion exchanges in the irreducible diagrams -- or potential --
figuring at NLO is found to be important in enlarging the domain of
validity of the effective field theory. We are able to reproduce
the scattering phase shift data up to $p=300$ MeV, with an
agreement which is superior to results of other effective field
theory approaches. We also discuss 
the importance of the explicit
pion degree of freedom in scattering process.

\vskip 1cm
\noindent
{\it PACS}: 03.65.Nk; 13.75.Cs; 11.15.Bt.\\
{\it Keywords}: Effective field theory; Neutron-proton scattering;
Finite cutoff scheme.
\vfill
\newpage

\renewcommand{\thefootnote}{\#\arabic{footnote}}
\setcounter{footnote}{0}

In a series of recent publications, two of the authors (DPM and
TSP) together with Kubodera and Rho~\cite{pkmr98,pkmrpp,pkmrnp} 
presented quite successful examples of effective field theory
in nuclear physics,
by showing response functions to electroweak processes 
at low energy involving two-nucleon systems
can be described quite accurately 
with little cutoff dependence.
(For others' works on this area, see
Refs. \cite{weinEW,Be95,Be97,Be99,vK98}.)
The approach used there was the ``$\Lambda$ counting scheme" proposed by
Weinberg~\cite{wein}.
The focus on the response functions instead of
on scattering observables is in the spirit of the long-standing
tradition in nuclear physics where the wealth of nuclear dynamics
has been accumulated more powerfully through response functions
than through scattering processes per se.

The recent surge of activity in effective field theories in nuclear
physics~\cite{INT-cal} was however triggered by the effort to give
a first-principle description of the large scattering lengths in
nucleon-nucleon scattering, so a large portion of the recent
publications has been devoted to scattering amplitudes at low
energy. ``Q counting" scheme introduced by Kaplan, 
Savage and Wise~\cite{KSW} using power divergence subtraction (PDS)
exemplifies the present preoccupation of the workers in this 
field.\footnote{\protect
There is also a slightly different scheme which emphasizes the
unitarity and the relativistic formalism 
suggested by Lutz \cite{lutz}.}
While there appeared a few papers on the power of the $\Lambda$
scheme in scattering~\cite{kolck, cohen, steele}, an exhaustive 
confrontation of
the $\Lambda$ scheme developed for the response
functions~\cite{pkmr98} with scattering amplitudes has not been
performed. It is the purpose of this paper to provide the missing
information. We shall show that the scheme is as successful for
scattering as for response functions and in fact is even more
accurate than the $Q$ counting scheme.

The question we address is this: How does the $\Lambda$ scheme of
\cite{pkmr98} which is remarkably successful in postdicting and
predicting electroweak processes of two-nucleon systems
$p+p\rightarrow d+e^++\nu$ \cite{pkmrpp} and $n+p\rightarrow d+\gamma$
\cite{pkmrnp,pmrnp} fare with the two-nucleon scattering 
and to what nucleon
momentum and with what accuracy can one ``push" the scheme?

For {\it np} scattering to the next-to-leading order (NLO)
in the $\Lambda$ scheme with pions and nucleons, the
two-particle irreducible graphs that generate the potential for the system 
with which the Schr\"{o}dinger(or Lippman-Schwinger) equation is to be solved 
comprise of nucleonic contact interactions and pion-exchange potentials. 
The contact terms, the coefficients of which will be fit to 
empirical data at the zero momentum limit, represent the short-range part 
of the nuclear interaction. 
On the other hand, the pion exchange potentials
control the long range part of the nuclear interaction, so
unsurprisingly the explicit inclusion of the pion-exchange potential
generally increases the range over which the wave functions are
accurate. It was shown in \cite{pkmr98} that the absence of the
OPEP could barely be compensated by the leading-order contact term 
and that the contact interaction at the
next-to-leading order (NLO) -- when added to the LO interaction -- is seen to
improve considerably the stability of the numerical
results, thereby increasing the domain of validity of the effective field
theory. In this paper, we will take explicitly into consideration 
the two-pion-exchange potential(TPEP) at the NLO and
show how accurately one can calculate the two-nucleon process.
In addition, we will also report on a higher order calculation.

We have described in great detail the main strategy of the cut-off EFT in our
previous works, so we will not enter into it here.
It will suffice to briefly define the convention and go directly
into the results.
We adopt the power counting rule given in \cite{wein}:
an irreducible diagram is of order of ${\cal O}((Q/\Lambda_\chi)^\nu)$
where $Q$ is the pion mass or the typical momentum scale of the process
and $\Lambda_\chi\sim 1$~GeV.
And we keep only pions and nucleons as pertinent degrees of freedom,
all other massive degrees of freedom are integrated out.
In nucleon-nucleon potential, the LO has $\nu=0$ and consists
of OPEP plus non-derivative contact interactions:
\begin{eqnarray}
{\cal V}_{LO}(\vec{q})= \frac{4\pi}{M} C_0 + {\cal
V}_{1\pi}(\vec{q}),\ \ \ {\cal V}_{1\pi}(\vec{q}) = \frac{g^2_A}{4
f^2_\pi} \frac{ {\vec q}^2}{m^2_\pi + \vec{q}^2}, \label{eq:lopot}
\end{eqnarray}
\noindent where $M$ is the nucleon mass, $g_A\simeq 1.25$ the
axial-vector coupling, $f_\pi\simeq 93\ \mbox{MeV}$ the pion decay
constant and $\vec{q}$ the momentum transfer. 
There is no $\nu=1$ contribution, the NLO ($\nu=2$)
comes from TPEP and contact interactions with two derivatives.
TPEP contains also $\nu=3$ (and higher order) contributions,
which correspond to the subleading order $\pi\pi NN$ vertices.
There have been some elaborate works on 
the TPEP \cite{rr96, kbw97, rtfs99} up to $\nu=3$ order
and applications of the potential in other channels as well as
the $S$-wave $np$ scattering.
In this work, we will not consider the $\nu=3$ order contributions,
though it may results in substantial improvement of the theoretical predictions.

\begin{figure}[tbp]
\centerline{ \setlength{\epsfxsize}{13cm}\epsffile{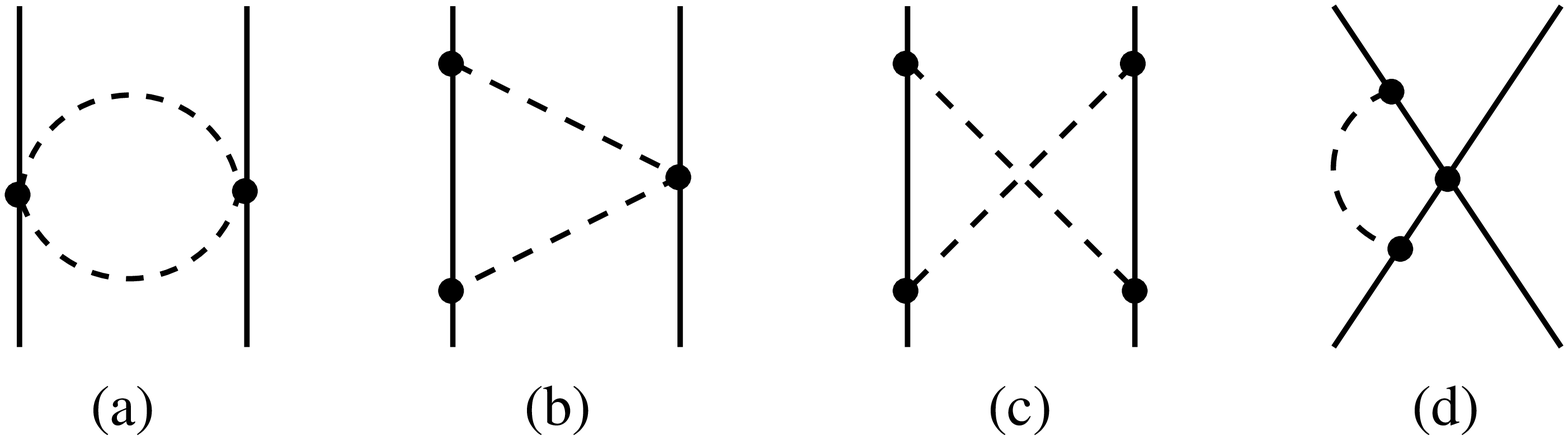} }
\caption[]{Two-body irreducible diagrams that contribute
to the TPE potential.}\label{fig:tpepfig}
\end{figure}

The {\it two-nucleon irreducible} diagrams that contribute to
the TPEP are shown in Fig.~\ref{fig:tpepfig}.
In addition, we also include the contributions from
OPEP-subtracted two-pion-box diagrams,
whose effects are negligible in the ${}^1S_0$ channel.
Remembering that the contact interaction with the coefficient $C_2$
enters at the same order, we write the full NLO potential
as
\begin{eqnarray}
{\cal V}_{\rm NLO}(\vec{q})=\frac{4\pi}{M}\ C_2\ \vec{q}^2 + {\cal
V}_{2\pi}(\vec{q}),
\end{eqnarray}
with
\be
{\cal V}_{2\pi}(\vec q)
 = \frac{1}{64\pi^2 f_\pi^4}\,\int_0^1\!dx\,
   \frac{({\vec q}^2)^2}{\frac{4 m_\pi^2}{1-x^2} + {\vec q}^2}
   \, f_{2\pi}(x)
\label{tpep}
\ee
\begin{eqnarray}
f_{2\pi}(x) = \frac{
% 4 x^4 + 8 g_A^2 x^2 (3 + 2 x^2) - 5 g_A^4 (3 + 36 x^2 + 8 x^4)}{24 (1 - x^2)},
 4 x^4 + 8 g_A^2 x^2 (3 + 2 x^2) - 4 g_A^4 (3 + 48 x^2 + 8 x^4)}{24 (1 - x^2)}.
\end{eqnarray} 
To the same order in
the chiral counting, there are loop corrections in the propagators and the
$\pi NN$ vertex. However in the kinematics involved in the elastic scattering,
they do not bring in any momentum-dependent corrections: They modify
only the constants (masses and coupling constants) which are to be absorbed
into the parameters extracted from experiments. \footnote{ \protect There 
are also certain contributions that come from relativistic corrections.
A detailed analysis of
these contributions is recently given by Friar \cite{friar}. Here
we are not concerned with them.} 

So far, we have assumed that $|\vec q|$
is of the same size of the pion mass $m_\pi$.
When $|\vec q|$ is much smaller than $m_\pi$,
we can integrate out even the pions.
Theory then contains only nucleons explicitly,
and the expansion parameter becomes $|\vec q|/m_\pi$, 
rather than $Q/\Lambda_\chi$.
As we see, OPEP and TPEP become of order of $|\vec q|^2/m_\pi^2$
and $|\vec q|^4/(2 m_\pi)^4$,
so they can be replaced by
suitable series of contact interactions starting from
two-derivatives and four-derivatives.
By the same token, multi-pion-exchange potentials 
can also be replaced by contact interactions.
Recalling that multi-pion-exchange potentials are hard to calculate,
this procedure provides us an efficient way of going to higher order.
The price of this simplification is the shrinking-down of
the radius of convergence.
Following to the general argument
of the effective-range expansion, the radius of convergence
is about a half of the lightest mass
of the degree of freedom that is not explicitly taken.
That is, without the pion degrees of freedom,
we can describe only up to $p \lsim m_\pi/2$,
where $p$ is the CM momentum.
Taking OPEP explicitly with higher order
contact interactions,
we then expect that we can achieve accurate description of the
scattering data up to $p \lsim m_\pi$,
which was indeed proved in \cite{pkmr98}.
In this paper we go to one further step,
by exploring the consequences of the inclusion of the TPEP.
With the TPEP, we expect we can go up to $p \lsim \frac32 m_\pi$.
Basically, in this work, we will focus on the
comparison between the NLO potentials with and without the TPEP. For 
clarity, we shall refer to the latter as the 
$1\pi C_2$, meaning that it contains OPEP and contact 
interactions up to $C_2$ term.
Thus in this notation, LO is identical to $1\pi C_0$ and
NLO is identical to $2\pi C_2$. 
One of our principal findings from this comparison between  
the NLO and $1\pi C_2$ potentials is that the explicit presence of the pion 
within the $\Lambda$ scheme makes the EFT a lot more versatile and 
accurate. In fact, this work provides a confirmation
of the partial finding in \cite{pkmr98} of the effect of the explicit OPEP 
in NLO and $1\pi C_2$ in scattering consideration.
We will also consider $2\pi C_4$ potential, which consists of
the NLO potential and four-derivative contact interaction
of the form $\frac{4\pi}{M} C_4 |\vec q|^4$.
This $2\pi C_4$ potential as a possible next 
subleading order may only find its justification for $|\vec q|$
smaller than $m_\pi$, whereas for other kinematical zone, 
the inclusion of all subleading NNLO TPEP should be made 
\cite{epelbaum}.
Thus compared to the NLO,
we expect that
the theory can improve (both in accuracy and in
cutoff-independence) results at that low momentum region, 
but not the radius of convergence in general.
These hypotheses will be proved in our work.
For the low momentum region with $p \lsim \frac32 m_\pi$,
we expect this potential provides us 
a simple but sufficiently adequate way to do an 
${\cal O}(|{\vec q}|^4)$-order
calculation without including pion two-loops.
It is worth noting that the integral in Eq.(\ref{tpep}) can be
rewritten in a spectral representation by changing the
integration variable $x$ to a ``spectral mass" $m \equiv \frac{2
m_\pi}{\sqrt{1-x^2}}$, which runs over $2 m_\pi \le m < \infty$. 
One may then introduce an upper limit to the spectral mass $m \le m_0$,
by imposing the range of the integration variable to be $0 \le x \le
\sqrt{1 - \frac{4 m_\pi^2}{m_0^2}}$. The $m_0$ should not be
confused with the genuine cutoff of the theory $\Lambda$ that
enters through the Fourier transformation as defined below. 
Note that the TPEP is negative while OPEP is positive for the whole
range of $|\vec{q}|$ and that it is quite small compared to the OPEP, less
than 3 \% for the momentum range $|\vec q| \le m_\pi$. This implies that
the results obtained in \cite{pkmr98} will not be changed appreciably by the
explicit inclusion of the TPEP. As the momentum increases, however,
the genuine TPEP will play an important role for the
radius of the convergence, the cutoff-dependence and the allowed
range of the cutoff of the theory. Throughout this work, we put
the value of $m_0$ large enough, $m_0=2\ \mbox{GeV}$ -- instead of
infinity -- to economize the calculation.

These potentials of various chiral orders, transformed into coordinate
space, will be put into the Schr\"{o}dinger equation. In doing the
Fourier transform, a cutoff in momentum space will be introduced. 
The potential with the cutoff $\Lambda$ takes the form
\begin{eqnarray}
V(\vec{r}) \equiv \int \frac{d^3 \vec{q}}{(2 \pi)^3}\ e^{i \vec{q}
\cdot\vec{r}}\ S_\Lambda(\vec{q}^2) \ {\cal V}(\vec{q})
\end{eqnarray}
with
\be
S_\Lambda(\vec{q}^2) = {\mbox e}^{- \frac{{\vec q}^2}{2
\Lambda^2}}. \ee 
Due to the cutoff -- which is kept finite, the integrals are finite so 
that we don't need counter terms in a proper sense
in addition to the contact terms that figure to
the order considered.  However, the concept of renormalization is still valid
and figures in the requirement that the results be stable against 
the variation of the cutoff. 
The Fourier transformation of the OPEP and the
contact terms can be done straightforwardly (see, for example,
\cite{pkmr98,lepage}). For the TPEP, a similar technique can be used
by performing the Fourier transformation before the parametric
space integration. The wave function $\psi_0(r) \equiv
\frac{u_0(r)}{r}$ is then obtained through the Schr\"{o}dinger
equation
\begin{eqnarray}
\left[ \frac{d^2}{dr^2} + M \left(E-V(r)\right)\right] u_0(r)=0
\end{eqnarray}
where $E = p^2/M$ is the total energy of the system, and
$p=|\vec{p}|$ is the center-of-mass momentum. The solutions are to
fit the {\it empirical} effective range expansion,
\begin{eqnarray}
p \cot \delta = - \frac{1}{a_0} + \frac{1}{2} r_e p^2 + v_2 p^4 +
{\cal O}(p^6). \label{eq:efrng}
\end{eqnarray}
For this we shall take the Nijmegen partial wave analysis (NPWA) \cite{skrs93}
as the empirical data and extract from it  
the values of the low-energy constants, $a_0=-27.73\ \fm$
and $r_e=2.677\ \fm$.   Note that the NPWA results are in very
good agreement with the values obtained by the Argonne $v_{18}$
potential\cite{v18}, i.e., $a_0(v_{18}) = -27.732\ \fm$ and $r_e(v_{18})
=2.697\ \fm$. The constant $C_4$ will be determined to reproduce
the effective volume $v_2=-0.48\ \fm^3$ \cite{hansen}, which was
extracted from the
NPWA data \cite{skrs93}.\footnote{\protect
In \cite{steele}, a slightly different renormalization procedure
has been introduced.
They are using the modified effective range expansion,
which deals with the OPEP-subtracted $NN$ potential.
Their method is not adopted in our work.}
We are now in position to discuss our results.
\begin{table}[t]
\begin{center}
\begin{tabular}{|c|c|r|r|r|r|r|}\hline
\multicolumn{2}{|c|}{$\Lambda$(MeV)}
&200&300&400&500&600 \\ \hline\hline
LO& $C_0$ &         $-1.27$ & $-0.83$ &$-0.62$ & $-0.50$ & $-0.42$ \\ 
\hline\hline
$1\pi C_2$ & $C_0$& $-1.32$ & $-1.21$ & $-0.35$ & 9.17 & 98.13 \\ \cline{2-7}
& $C_2 \Lambda^2$&  0.06 & 0.70 & 1.77 & 5.54 & 23.52 \\ \hline\hline
NLO & $C_0$ &       $-1.31$&$-1.24$&$-1.08$&0.14&11.43 \\ \cline{2-7}
 & $C_2 \Lambda^2$& 0.11&0.85 &1.74 &3.27& 7.48 \\ \hline\hline
$2\pi C_4$ & $C_0$& $-1.32$& $-0.68$ & 16.40 & & \\ \cline{2-7}
 & $C_2 \Lambda^2$& $-0.12$&0.75 & 7.12 & & \\ \cline{2-7}
 & $C_4 \Lambda^4$& 0.14& 0.38 & 1.56 & & \\ \hline
\end{tabular}
\caption[]{ 
Coefficients of the contact interaction terms in
LO, $1\pi C_2$, NLO and $2\pi C_4$ potentials,
for different values of the cutoff $\Lambda$.
We have multiplied $\Lambda^n$ to the $C_n$ ($n=0,\, 2,\,4$)
so that the unit of the $C_n \Lambda^n$ is $fm$ for any $n$.
The blanks for the $2\pi C_4$ are explained in the text.
} \label{tab:cons}
\end{center}
\end{table} 

In confronting the experiment with our theory, we shall particularly be 
interested in assessing how well the basic tenets 
of an effective field theory of the type we are considering 
are satisfied \cite{lepage}. The first is that
the effective field theory
in question is a nonrenormalizable theory and a truncated
version with a given cutoff scale must break down at some point signaling the
emergence of ``new physics." The second point is that the effective theory with
a cutoff regularization is valid only if the physical observables calculated
within the scheme 
{\em become stable} against the variation of the cutoff 
{\em when higher order terms are included}, which is 
nothing but an approximate manifestation of the renormalization group (RG)
invariance that any valid effective theory must satisfy.

\begin{figure}[tbp]
%\centerline{ \setlength{\epsfxsize}{13cm}\epsffile{p140-com.ps} }
%\centerline{ \setlength{\epsfxsize}{13cm}\epsffile{p280-com.ps} }
\centerline{ \setlength{\epsfxsize}{13cm}\epsffile{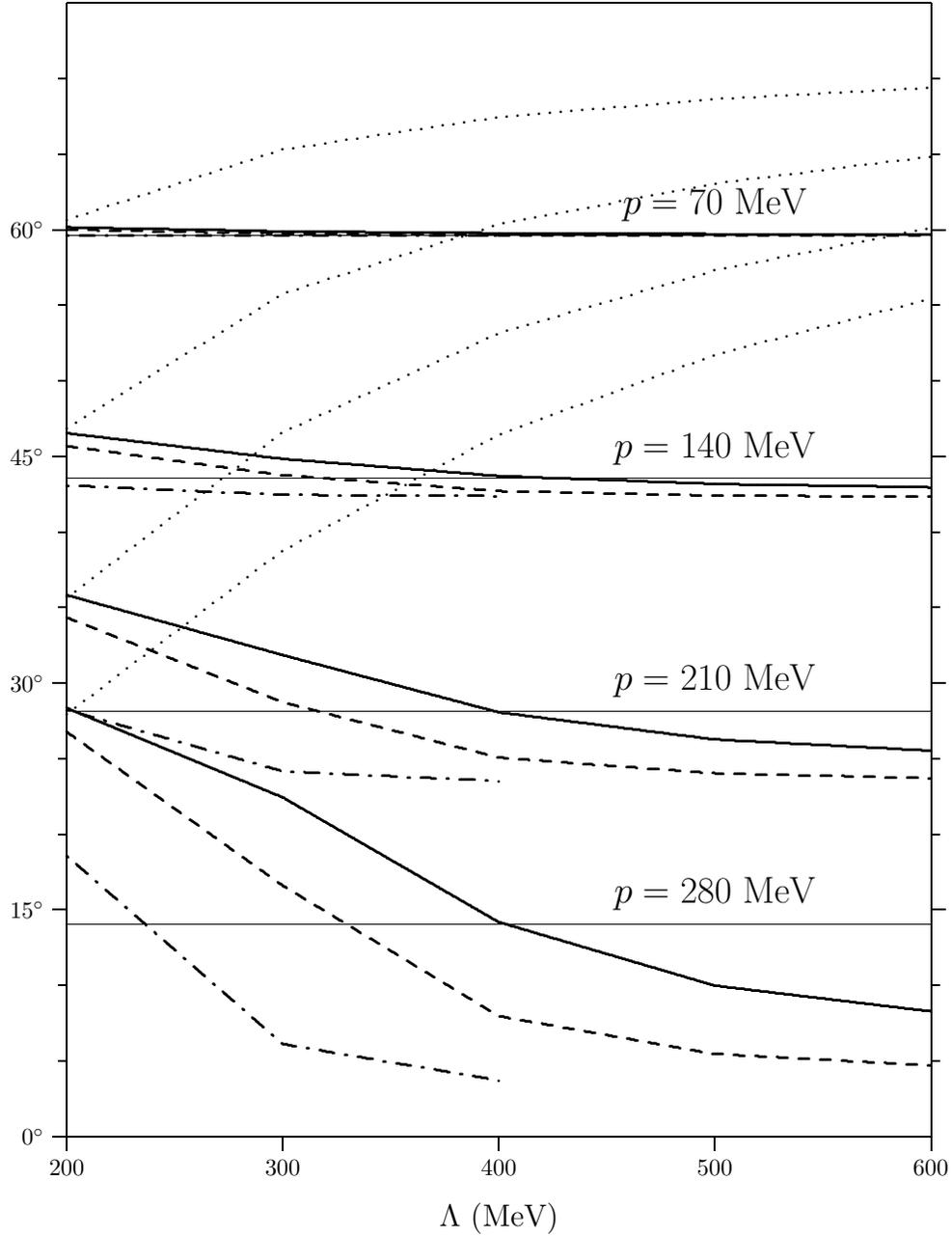} }
\caption[]{Phase shift with respect to $\Lambda$ 
at $p=70$, 140, 210 and 280 MeV.
For each momentum, the LO, $1\pi C_2$, NLO and $2\pi C_4$ results are 
represented by dotted, dashed, solid and dotted-dashed curves.
}\label{fig:p280/140}
\end{figure}

Let us see how these points emerge in the calculation. 
In Table~\ref{tab:cons} are displayed the values of
the coefficients for various potentials.
We can observe in the Table that, except for the LO case,
the values of $C$'s increase rapidly at a certain value of the cutoff.
In fact there is a critical value of the cutoff
beyond which we cannot find the $C$'s that reproduces the low-energy
constants \cite{cohenere}.
The $2\pi C_4$ for $\Lambda \geq 500$ MeV is found to be
such a case.\footnote{
If we allow a slight deviation by about $0.03\ {\mbox{fm}}^3$
in the effective range,
we can go at least up to $\Lambda=600$ MeV.}
We show in Fig.~\ref{fig:p280/140} the $^1$S$_0$ phase shift as
a function of the cutoff $\Lambda$ for the initial CM momenta of
70, 140, 210 and 280 MeV (from top to bottom) for different potentials. 
In the figure, 
the NPWA data are represented by thin horizontal solid lines;
the LO, $1\pi C_2$, NLO and $2\pi C_4$ results by dotted, 
dashed, solid and dotted-dashed curves, respectively.
The extensive comparison between LO (dotted curves) and 
$1\pi C_2$ (dashed curves)
in this framework has been given in \cite{pkmr98}:
By adding $C_2$ term into the LO potential,
the $1\pi C_2$ theory becomes much more accurate with much less
cutoff dependence compared to the LO theory.
Indeed, even at $p=140$ MeV where the theory is
expected to break down, the theory
differs from experiment only by less 
than 3 \% for all the region with $\Lambda \geq$ 300 MeV. 
We also observe that the theory becomes closest to the
experiment at $\Lambda\sim 300$ MeV,
while the LO favors $\Lambda \sim 200$ MeV.
Now by adding the TPEP, we see that
the NLO (solid curves) becomes more accurate
than the $1\pi C_2$ at higher cutoff.
For all the region considered with $\Lambda \geq 300\ \mbox{MeV}$,
the NLO differs from experiment less than 13 \%
at $p=210\ \mbox{MeV}$.
Recalling that the theory is expected to break down
for $p \gsim \frac32 m_\pi$,
the above result is encouraging.
It is also worthwhile observing that
the NLO with $\Lambda=400$ MeV nearly coincides with
the NPWA 
for all the momentum considered, $p \leq 280$ MeV.
On the other hand the improvement between the $1\pi C_2$ and the NLO
is not spectacular.
The fact that the $2\pi C_4$ cannot come closer to the NPWA
shows the importance of the omitted
subleading TPEP and higher order terms.
%Indeed, the fact
%that the $2\pi C_4$ cannot come closer to the NPWA is a 
%clear illustration that the cut-off EFT is failing. 
%The missing parts, i.e., 
%the subleading TPEP and higher order terms,
%must account for the instability of the calculation.
Thus mere inclusion of higher-order contact interactions 
cannot possibly ameliorate the convergence.  
This indicates that any attempt  
to improve the precision by merely including higher
order contact terms 
without the explicit account of the relevant degrees of freedom
that are nearby is doomed to fail.
%{\tt Either:}
%This observation should be a general remark,
%not limited to the cut-off EFT.
%{\tt Or:}
%This observation does not
%necessarily indicate the failure of the cut-off EFT, 
%only that of a
%particular partial calculation within that effective theory. 

In the figure, it is clearly seen that the theories with higher terms
become more insensitive to the cutoff,
which shows the realization of the above mentioned RG invariance.
At $p=280$ MeV, while the NLO shows reasonable agreement with experiment,
results show large fluctuations with respect to
cutoff which implicates the breakdown of the theory. 
{}From Fig. 3, one can confirm that the theory breaks down at
around this momentum.

For completeness, we shall now compare our cutoff scheme with the PDS
scheme\cite{Rupak,pdsnp}. 
\begin{figure}[t]
\centerline{ \setlength{\epsfxsize}{12cm}\epsffile{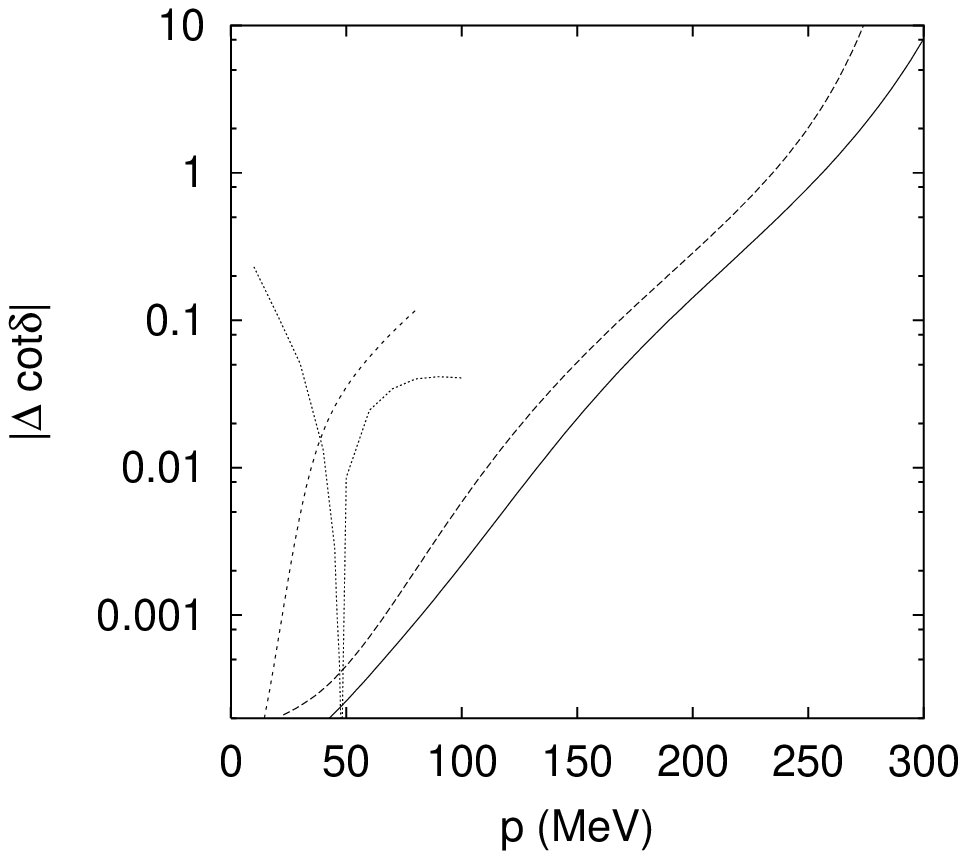} }
\caption[]{$|\Delta \cot\delta|(\equiv  |\cot\delta_{\mbox{\small
EFT}} - \cot\delta_{\mbox{\small NPWA}}|$) with respect to the
momentum $p$. Solid line is for NLO($=2\pi C_2$) and
dashed line for $1\pi C_2$. 
%and short dashed line for $2\pi C_4$.
Dotted line is the results of PDS fitted to $a_0$ and $r_e$ and 
dot-dashed line is the results for fit over $p\leq 200$ MeV.} 
\label{fig:comp}
\end{figure}
For ease of comparison, we estimate the deviation from the NPWA result for
the effective range function $\cot\delta$ of the cutoff EFT and PDS schemes
as a function of the momentum.
The results are shown in Fig. \ref{fig:comp}. Our
NLO results are drawn in solid curve
and the $1\pi C_2$ in dashed curve.
% and the $2\pi C_4$ in short dashed curve. 
The PDS results
obtained up to NLO are drawn both for the coefficients 
fit to the low-energy constants, $a_0$ and $r_e$ (dotted line)
\cite{cohen} and for the results obtained by a global fit for the
range $p \le 200\ \MeV$ (dot-dashed line). 
%It might be worthy to note that
%the NLO in PDS is suppressed by $Q$,
%while the NLO in our framework is suppressed by $Q^2$.
%However, both theories have the same number of free parameters.
It might be worthy to note that
the NLO in PDS and in our framework have different meanings,
as the counting rules adopted are different.
Nevertheless both theories have the same number of free parameters,
so the comparison can be meaningful.
{}From the
figure, one can confirm that the cutoff theory is doing better than the
PDS results in accuracy. A similar observation was made in
\cite{cohen}. 
Recently, the NNLO calculation in the PDS method 
(which is $Q^2$ order compared to the LO) was reported,
both in a toy model \cite{Rupak} and in a real situation \cite{Flem}.
In \cite{Flem}, they have a good agreement up to $p=400$ MeV
by fitting the parameters over $p=7 - 200$ MeV.
Due to their fitting procedure, it is however meaningless
to directly compare their results to ours.\footnote{
As a consequences of the fitting, their low-energy constants
are different from experiment. For example,
the effective volume in their approach is $v_2=-1.2\ {\mbox{fm}}^3$,
which should be compared to the data $-0.48\ {\mbox{fm}}^3$.}
In applying their framework to the spin-one channels,
they conclude that the nonperturbative treatment of potential
is needed.

To conclude,
we have demonstrated in this letter that 
the $\Lambda$ scheme can be made markedly more successful 
for the scattering problem by going to the NLO order {\it a la } 
Weinberg. We have also shown the importance of the explicit presence of 
the pion exchange potential in NLO, extending the domain of validity as 
well as increasing the precision. The method is found to be
as successful in scattering as in electroweak responses.  
It has also been discussed that there are missing parts in the TPEP,
the subleading order contributions,
which are expected to play a significant role in increasing the accuracy at
high momentum region.
While we are revising this paper,
quite a similar work has been reported
by Epelbaum, Gl\"ockle and Mei{\ss}ner \cite{epelbaum}.
They have constructed the potential
including up to the subleading TPEP
(and the leading TPEP but with also Delta isobar).
They have performed the calculation in momentum space,
and applied to various channels as well as the ${}1S_0$ $np$ scattering.
And they have got similar conclusions to ours,
including the importance of the subleading TPEP.

We are particularly grateful to Mannque Rho for his continuous supports
and invaluable discussions during this work. Work of DPM and CHH is 
partially supported by the KOSEF through CTP of SNU and by the Korea
Ministry of Education under contract No. 99-2418 and BK-21.


\thebibliography{99}
\bibitem{pkmr98} T.-S. Park, K. Kubodera, D.-P. Min and M. Rho,
   Phys. Rev. {\bf C58} (1998) 637;
   Nucl. Phys. {\bf A646} (1999) 83;
   T.-S. Park, hep-ph/9803417.

\bibitem{pkmrpp}  T.-S. Park, K. Kubodera, D.-P. Min, M. Rho,
Astrophys. Jour. {\bf507} (1998) 443.

\bibitem{pkmrnp} T.-S. Park, K. Kubodera, D.-P. Min and M. Rho, 
nucl-th/9904053; nucl-th/9906005.


\bibitem{weinEW} S. Weinberg, Phys. Lett. {\bf B295} (1992) 114.

\bibitem{Be95}
S.~R. Beane, C.~Y. Lee and U.~van Kolck, Phys. Rev. {\bf C52} (1995) 2914.

\bibitem{Be97}
S.R. Beane, V.~Bernard, T.-S.~H. Lee, U.-G.~Mei{\ss}ner and U.~van Kolck,
  Nucl. Phys. {\bf A618} (1997) 381.

\bibitem{Be99}
S.R. Beane, M.~Malheiro, D.R. Phillips and U.~van Kolck, nucl-th/9905023.

\bibitem{vK98}
U.~van Kolck, Nucl. Phys. {\bf A645} (1999) 273.


\bibitem{wein} S. Weinberg, Phys. Lett. {\bf B251} (1990) 288;
 Nucl. Phys. {\bf B363} (1991) 3; Phys. Lett. {\bf B295} (1992) 114.

\bibitem{INT-cal} INT-Caltech 1998 workshop proceedings, {\it Nuclear Physics
with Effective Field Theory}, ed. R. Seki, U. van Kolck, and M.J. Savage,
World Scientific, 1998

\bibitem{KSW} D.B. Kaplan, M.J. Savage and M.B. Wise,
 Phys. Lett. {\bf B424}, 390 (1998).

\bibitem{lutz} M. Lutz, nucl-th/9906028.

\bibitem{kolck} C. Ord\'{o}\~{n}ez and U. van Kolck, 
Phys. Lett. {\bf B291} (1992) 459; 
U. van Kolck, Texas Ph.D. Dissertation (1993);
C. Ord\'{o}\~{n}ez, L. Ray and U. van Kolck, Phys. Rev. Lett. {\bf 72} 
(1994) 1982; Phys. Rev. {\bf C53} (1996) 2086. 

\bibitem{cohen} T.D. Cohen and J.M. Hansen,
Phys. Rev. {\bf C59} (1999) 13;
 D.R. Phillips and T.D. Cohen, nucl-th/9906091.

\bibitem{steele} J. V. Steele and R. J. Furnstahl,
Nucl. Phys. {\bf A645} (1999) 439.

\bibitem{pmrnp} T.-S. Park, D.-P. Min and M. Rho, 
Phys. Rev. Lett. {\bf 74 } (1995) 4153;
Nucl. Phys. {\bf A596} (1996) 515.

\bibitem{rr96} M. Robilotta and da Rocha, Nucl. Phys.
{\bf A615} (1996) 2086.

\bibitem{kbw97} N. Kaiser, R. Brockmann and W. Weise,
Nucl. Phys. {\bf A625} (1997) 758.

\bibitem{rtfs99} M. C. M. Rentmeester, R. G. E. Timmermans,
J. L. Friar and J. J. de Swart,Phys. Rev. Lett. {\bf 82} (1999) 4992.

\bibitem{friar} J.L. Friar, nucl-th/9901082; U. van Kolck, nucl-th/9902015.

\bibitem{epelbaum}
E. Epelbaum, W. Gl\"ockle and Ulf-G. Mei{\ss}ner, nucl-th/9910064.

\bibitem{lepage} G.P. Lepage, nucl-th/9706029; talk given
at the Workshop on ``Nuclear
Physics with Effective Field Theories," INT, University of
Washington, February 25-26, 1999.

\bibitem{skrs93} V.G.J. Stoks, R.A.M. Klomp, M.C.M. Rentmeester and J.J. de
Swart, Phys. Rev. {\bf C48} (1993) 792.

\bibitem{v18} R. B. Wiringa, V. G. J. Stoks and R. Schiavilla,
Phys. Rev. {\bf C51} (1995) 38.

\bibitem{hansen} T.D. Cohen and J.M. Hansen,
Phys. Rev. {\bf C59} (1999) 13.

\bibitem{cohenere} S.R. Beane, T.D. Cohen and D.R. Phillips,
Nucl. Phys. {\bf A632} (1998) 445;
K.A. Scaldeferri, D.R. Phillips, C.-W. Kao and T.D. Cohen,
Phys. Rev. {\bf C56} (1997) 679; 
Daniel R. Phillips and Thomas D. Cohen,
Phys. Lett. {\bf B390} (1997) 7. 

\bibitem{Rupak} G. Rupak and N. Shoresh, nucl-th/9902077.

\bibitem{pdsnp}  J.-W. Chen, G. Rupak, M.J. Savage, nucl-th/9905002.

\bibitem{Flem} S. Fleming, T. Mehen and I.W. Stewart, nucl-th/9911001.
\end{document}